\documentstyle[prl,aps,twocolumn,amssymb,graphicx]{revtex}
\def\be{\begin{equation}}
\def\ee{\end{equation}}
\def\bea{\begin{eqnarray}}
\def\eea{\end{eqnarray}}
\def\gtap{\ \raisebox{-.4ex}{\rlap{$\sim$}} \raisebox{.4ex}{$>$}\ }
\def\ltap{\mathrel{
   \rlap{\raise 0.511ex \hbox{$<$}}{\lower 0.511ex \hbox{$\sim$}}}}

\newcommand{\meff}{\mbox{$\langle m_\nu \rangle$}}

\newcommand{\deltaatm}{\mbox{$\Delta m^2_{\oplus}$}}
\newcommand{\deltasol}{\mbox{$ \Delta m^2_{\odot}$}}

\newcommand\lsim{\mathrel{\rlap{\lower4pt\hbox{\hskip1pt$\sim$}}
    \raise1pt\hbox{$<$}}}
\newcommand\gsim{\mathrel{\rlap{\lower4pt\hbox{\hskip1pt$\sim$}}
    \raise1pt\hbox{$>$}}}

\def\Jo#1#2#3#4{{\it #1} {\bf #2}, #3 (#4)}

\def\circa#1{\,\raise.3ex\hbox{$#1$\kern-.75em\lower1ex\hbox{$\sim$}}\,}

\def\NPB{{\rm Nucl. Phys.} {\bf B}}

\def\PLB{{\rm Phys. Lett.}  {\bf B}}
\def\PRL{\rm Phys. Rev. Lett.}
\def\PRD{{\rm Phys. Rev.} {\bf D}}

\begin{document}
\preprint{} 
\draft

%
%
\input epsf
\renewcommand{\topfraction}{0.99}
\renewcommand{\bottomfraction}{0.99}
\twocolumn[\hsize\textwidth\columnwidth\hsize\csname@twocolumnfalse\endcsname

\title{Connecting Low Energy Leptonic CP-violation  to  Leptogenesis}
\author{S. Pascoli$^{1,2}$, S.T. Petcov$^{3}$ and Antonio Riotto$^{1,4}$}

\address{$^{1}$CERN, Theory Division, Geneve 23, CH-1211 Switzerland}
\address{$^{2}$IPPP, Dept. of Physics, University of Durham, DH1 3LE,UK}
\address{$^{3}$SISSA and INFN, Sezione di Trieste, I-34014, Trieste, Italy}
\address{$^{4}$INFN, Sezione di Padova, Via Marzolo 8, 35131, Italy}
\date{\today}
\maketitle
\begin{abstract}
\noindent
It was  commonly  thought   
that the observation of 
low energy leptonic CP-violating phases     would not automatically 
imply the existence of a baryon asymmetry in the leptogenesis scenario. 
This conclusion does not generically hold  when 
the issue of flavour is relevant and 
properly taken into account in leptogenesis. 
We illustrate this point with  various 
examples studying the correlation  between the baryon asymmetry and 
the CP-violating asymmetry 
 in neutrino oscillations  and  the  effective Majorana mass
in neutrinoless double beta decay.
\end{abstract}


\vskip2pc]


\noindent
Leptogenesis \cite{FY} is 
a simple  mechanism to 
explain the baryon number asymmetry (per entropy density)
of the Universe  $Y_{\cal B}=\left(0.87\pm 0.02\right)
\times 10^{-10}$ \cite{wmap}.
A lepton asymmetry is dynamically generated
and then  converted into a baryon asymmetry
due to $(B+L)$-violating sphaleron interactions \cite{kuzmin,baureview}
which exist in the Standard Model (SM).
A simple model in which this mechanism can be implemented is the 
``seesaw''(type I)
\cite{seesaw}, consisting
of the SM plus three   right-handed (RH) Majorana neutrinos.
In thermal leptogenesis \cite{leptogen} the heavy RH neutrinos
are produced by thermal scatterings after inflation  and subsequently
decay out-of-equilibrium in a lepton number and CP-violating way, 
thus satisfying Sakharov's constraints \cite{baureview}.
At the same time the smallness of  neutrino masses
suggested by oscillation experiments \cite{exp}
can be ascribed to the seesaw mechanism where integrating out 
heavy RH Majorana neutrinos  generates mass terms for the left-handed 
flavour neutrinos
which are inversely proportional to the mass of the RH ones.

Establishing a connection between the CP-violation in low energy neutrino 
physics and the CP-violation at high energy necessary for leptogenesis
has received much attention in recent years 
\cite{comb}
and is the subject of the present paper. 
In the case of three neutrino mixing, 
CP-violation at low energy is parameterized by the  
phases in the 
Pontecorvo--Maki--Nagakawa--Sakata (PMNS) 
\cite{PMNS} lepton mixing matrix $U$. It contains the 
Dirac phase
$\delta$ and, if neutrinos are Majorana particles, 
two Majorana phases $\alpha_{21}$ and $\alpha_{31}$
\cite{majorana}.
The Dirac phase $\delta$ enters in the probability of neutrino
oscillations. 
The corresponding CP-asymmetry is  given by
the difference between the oscillation probability 
for neutrino and antineutrinos, 
$\Delta P=P\left(\nu_\mu\rightarrow \nu_e\right)-
P\left(\overline{\nu}_\mu\rightarrow \overline{\nu}_e\right)\propto
J_{\rm CP}$ where the rephasing invariant 
$J_{\rm CP}={\rm Im}\left(U_{e1}U^*_{e2}U^*_{\mu 1}U_{\mu 2}\right)$
\cite{kp}
is proportional to $\sin 2 \theta_{13} \sin \delta$. This  
implies that the observation prospects of CP-violation 
in future long-baseline experiments
depend on the true value of 
$\sin 2\theta_{13}$. 
Present studies indicate that
a wide range of values of the $\delta$ phase
could be tested in 
superbeam and betabeam experiments
if $\sin^2 2 \theta_{13}\simeq {\rm few} \times (10^{-3}-10^{-2})$,
or in a future neutrino factory even if 
$\sin^2 2 \theta_{13}$ is as small as $10^{-4}$.
The two Majorana CP-violating phases enter
only processes at low energy in which the lepton
number is violated by two units.
The most sensitive of these processes is neutrinoless double beta decay,
which is currently under intensive experimental search \cite{aa}.
The decay rate is a function of  the effective Majorana mass
$\langle m_\nu\rangle=\left(m_1\,U_{e1}^2+m_2\,U_{e2}^2+m_3\,U_{e3}^2\right)$ 
which depends on the 
type of neutrino mass spectrum.
Typically, one can consider the 
normal hierarchical (NH) ($m^2_1 \ll m^2_2 \simeq \Delta m^2_\odot \ll
m^2_3 \simeq \Delta m^2_{\oplus}$),
inverted hierarchical (IH) ($m^2_3 \ll m^2_1 \simeq m^2_2 \simeq 
\Delta m^2_{\oplus}$),
and quasi-degenerate (QD)  ($m^2_1 \simeq m^2_2 \simeq m^2_3 \gtap
\Delta m^2_{\oplus}$) 
spectra. Here $\Delta m^2_{\odot}$ and $\Delta m^2_{\oplus}$ are the
mass square differences which drive  the solar and the atmospheric neutrino 
oscillations, 
respectively and $m_i$ ($i=1,2,3$) are the light neutrino masses.
One Majorana phase can, in principle, be observed 
 although this represents a challenge.
For a detailed discussion see Refs.~\cite{PPRCP,PPS}.


It was  commonly accepted  that
the future observation of leptonic low energy CP-violation 
would not automatically 
imply a nonvanishing baryon asymmetry through leptogenesis. 
This conclusion, however, was shown in 
\cite{davidsonetal,nardietal,davidsonetal2} not to hold universally.
The reason is based on  a new ingredient
recently accounted for in the leptogenesis scenario, 
lepton flavour \cite{davidsonetal,nardietal,davidsonetal2,barbieri}. 
The    dynamics of
leptogenesis is usually  addressed within the `one-flavour' approximation,
where  Boltzmann equations
are written for the abundance of the 
lightest RH neutrino 
and  for the total lepton asymmetry. However, this approximation is
rigorously correct only when the interactions mediated by 
charged lepton Yukawa couplings are out of equilibrium. Supposing that
leptogenesis takes place at temperatures $T\sim M_1$, where $M_1$
is the mass of the lightest RH neutrino, the `one-flavour' approximation 
only holds for $M_1\gsim 10^{12}$ GeV. In this range all
the interactions mediated by  the 
charged lepton Yukawa couplings are out of equilibrium and there is no notion
of flavour. One is allowed to perform a rotation
in flavour space to store all the lepton asymmetry in one flavour, the
total lepton number. 
However, at $T\sim M_1\sim 10^{12}$ GeV, the interactions mediated by the
charged tau Yukawa coupling  come into equilibrium followed by those mediated
by the charged muon Yukawa coupling at $T\sim M_1\sim 10^{9}$ GeV and the
notion of flavour becomes physical.
Including the issue of flavour can  significantly affect the result for the
final baryon asymmetry \cite{davidsonetal,nardietal,davidsonetal2}. 
Thermal leptogenesis is a dynamical
process, involving  the  production and 
destruction of RH neutrinos 
and  of the  lepton asymmetry  that is distributed among
distinguishable flavours.  The processes which
wash out lepton number
are flavour dependent, {\it e.g.} the inverse decays
from electrons can destroy the lepton asymmetry carried by,
and only by,  the electrons.
The  asymmetries  in each flavour
are  therefore washed out differently,
and will appear with different weights in the final formula
for the baryon asymmetry.  This is physically inequivalent
to the treatment of washout in the one-flavour approximation,
 where  
the flavours are taken indistinguishable, thus obtaining  the unphysical
result that 
 inverse decays  from all flavours are taken to wash out asymmetries
in any flavour (that is, {\it e.g.}, 
 an asymmetry stored in the first family may be washed
out by inverse decays involving the second or the third family). 

When flavour is accounted for, 
the final value of the baryon asymmetry is the sum of three
contributions. Each term is given by the 
CP asymmetry in a given flavour    $\alpha$ properly weighted
 by a washing  out factor induced by the    
lepton $\alpha$ violating processes.  
Taking into account the flavour 
dependence one may show that   observing low energy
CP-violating phases automatically implies, barring accidental  cancellations,  
generation of the
baryon asymmetry.
Before going into details though, let us summarize  
why this conclusion 
 is  not  possible
in the   `one-flavour' approximation.
The starting point is the Lagrangian of the SM
with the addition of three right-handed neutrinos $N_{i}$ ($i=1,2,3$)
with heavy Majorana masses $M_{i}$
and Yukawa couplings $\lambda_{i\alpha}$. Working in
the basis in which the Yukawa couplings for the
charged leptons are  diagonal, the Lagrangian reads
\begin{equation}
\label{L}
{\cal L} = {\cal L}_{\rm SM} + \frac{M_{i}}{2} N_i^2 + 
\lambda_{i\alpha} N_i  \ell_\alpha \, H +\hbox{h.c.}\, .
\end{equation}
Here $\ell_\alpha$ indicates the lepton doublet with flavour 
$(\alpha=e,\mu,\tau)$ and $H$ is the Higgs doublet whose vacuum expectation 
value is 
$v$. For the time being, we 
assume that right-handed neutrinos are hierarchical,
$M_{2,3}\gg M_1$ so that restricting to the dynamics of 
$N_1$ suffices. 

The total lepton asymmetry  per entropy density generated by the
$N_1$ decays is given by $Y_{\cal L}\simeq  
(\epsilon_1/g_*)\eta\left(\widetilde{m_1}\right)$,
where $\eta\left(\widetilde{m_1}\right)$ accounts for the 
washing out of the total lepton asymmetry due to $\Delta L=1$ inverse decays, 
$\widetilde{m_1}=(\lambda\lambda^\dagger)_{11}v^2/M_1$, $g_*$ counts the
relativistic degrees of freedom  and 
the CP asymmetry 
generated by $N_1$ decays  reads 
\begin{eqnarray}
\epsilon_{1} &\equiv& 
\frac{\sum_\alpha[\Gamma(N_1\rightarrow H \ell_\alpha)-\Gamma(
N_1\rightarrow \overline{H} \overline{\ell}_\alpha)]}{
\sum_\alpha[\Gamma(N_1\rightarrow H
\ell_\alpha)+\Gamma
(N_1\rightarrow \overline{H} \overline{\ell}_\alpha)]}\nonumber\\
&=&-
\frac{3 M_1}{16\pi}\sum_{j\neq 1}
\frac{\textrm{Im}
\left[ (\lambda \lambda^{\dagger})_{1 j}^2\right] }{\left[\lambda
\lambda^{\dagger}\right]_{11}}\frac{1}{M_j}\, .
\end{eqnarray}
Notice, in particular, that the CP asymmetry in the 'one flavour
approximation' depends  upon the trace of the CP asymmetries
over flavours. In the basis where the charged lepton Yukawa coupling
and the RH mass matrix are diagonal, the neutrino
Yukawa matrix can be written as $\lambda=V_R^\dagger{\rm Diag}(\lambda_1,
\lambda_2,\lambda_3)V_L$ and  the low energy leptonic phases may arise from the
phases in the left-handed (LH) sector, in RH sector, or from both.
The CP-asymmetry  can 
be expressed   in terms
of the diagonal matrix of the light neutrino mass eigenvalues $m={\rm 
Diag}(m_1,m_2,m_3)$, the diagonal matrix of the the right
handed neutrino masses $M={\rm Diag}(M_1,M_2,M_3)$ 
and an orthogonal complex matrix  
$R=v M^{-1/2}\lambda U m^{-1/2}$ \cite{Casas:2001sr}, 
which ensures that the correct low energy parameters 
are obtained. CP-violation in the RH sector is encoded in the
phases of $V_R$ and, from $\lambda\lambda^\dagger=V_R^\dagger{\rm Diag}(
\lambda_1^2,\lambda_2^2,\lambda_3^2)V_R=M^{1/2}R m R^\dagger M^{1/2}/v^2$, one
sees that the phases of $R$ are related to those of $V_R$. Now, 
summing over all flavours, one finds

\begin{equation}
\epsilon_{1}= 
-\frac{3 M_1}{16\pi v^2}\, \frac{{\rm Im}\left(
\sum_{\rho} m_\rho^{2} 
R^2_{1\rho}\right)}{\sum_\beta m_\beta\left|R_{1\beta}\right|^2}\, .
\end{equation}
In the `one-flavour' approximation a future observation of CP-violating
phases in the neutrino sector does not imply the existence
of a baryon asymmetry. Indeed, low
energy CP phases might stem  entirely from the LH sector and hence be
irrelevant for leptogenesis which would be  driven by the phases in $R$, {\it
i.e.} of the RH sector..

The  `one-flavour' approximation
rigorously holds, however, 
 only when the interactions mediated by the charged lepton
Yukawas are out of equilibrium, that is at $T\sim M_1\gsim 10^{12}$ GeV.
In this regime, flavours are indistinguishable and there is effectively 
only one flavour, the total lepton number.
At smaller temperatures, though, flavours are distinguishable: the
$\tau$ ($\mu$) lepton doublet is  a distinguishable mass eigenstate
for $T\sim M_1\lsim 10^{12}\,(10^9)$ GeV. 
The asymmetry
in each flavour is given by 
\begin{equation}
\label{q}
\epsilon_{\alpha}= -\frac{3 M_1}{16\pi v^2}\, \frac{{\rm Im}\left(
\sum_{\beta\rho} 
m_\beta^{1/2}m_\rho^{3/2} U^*_{\alpha\beta}U_{\alpha\rho}
R_{1\beta}R_{1\rho}\right)}{\sum_\beta m_\beta\left|R_{1\beta}\right|^2}\, .
\end{equation}
The trace over the flavours of  $\epsilon_{\alpha}$ 
coincides of course with $\epsilon_1$. 
Similarly, one may  define a parameter for each flavour 
$\alpha$, $\widetilde{m_{\alpha}}=|\lambda_{1\alpha}|^2 v^2/M_1$
parametrizing the  decay rate of $N_1$ to the $\alpha$-th 
flavour and   the trace  $\sum_\alpha \widetilde{m_{\alpha}}$
 coincides with the
$\widetilde{m_1}$ parameter defined  for the one-single flavour case.
Solving the Boltzmann equations for each flavour one finds
 $Y_{\alpha}\simeq
 (\epsilon_{\alpha}/g_*)\eta\left(\widetilde{m_\alpha}\right)$ 
\cite{davidsonetal,nardietal,davidsonetal2}.  
The way the total baryon asymmetry
depends upon the individual lepton asymmetries is a function
of  temperature. For instance, for 
$(10^9\lsim T\sim M_1\lsim 10^{12})$ GeV, only the interactions
mediated by the $\tau$ Yukawa coupling are in equilibrium and the 
final baryon asymmetry is $
Y_{\cal B}=-(12/37g_*)\left(\epsilon_2\eta\left(0.7\,
\widetilde{m_2}\right)+
\epsilon_\tau\eta\left(0.67\,\widetilde{m_\tau}\right)\right)$, 
where $\epsilon_2=\epsilon_{e}+\epsilon_{\mu}$, $\widetilde{m_2}=
\widetilde{m_e}+
\widetilde{m_\mu}$, $Y_{2}=Y_{e+\mu}$ \cite{davidsonetal2}.
As the CP asymmetry in each
flavour is weighted by the corresponding wash out parameter, 
$Y_{\cal B}$  is
generically not proportional  to $\epsilon_1$, 
but depends on each $\epsilon_\alpha$.
The dependence on the PMNS matrix elements
in (\ref{q})  is such that  
nonvanishing low energy leptonic CP-violating phases imply, in the context of
leptogenesis and barring accidental cancellations,  a nonvanishing  
baryon asymmetry  \cite{nardietal,davidsonetal2}. 

We  
can go even further. 
CP invariance would correspond to a real matrix $R$ provided that 
the CP-parities of 
the heavy and light Majorana neutrinos 
are equal to $+i$ \cite{appear}. In this case the low energy
Majorana phases vanish  (modulo $2\pi$) and $\delta =0$ (modulo $\pi$). 
$R$  real \cite{nardietal,davidsonetal2}
corresponds to the class
of models where CP is an exact symmetry in the
RH neutrino sector \cite{davidsonetal2}. 
In this case, the flavour CP asymmetries and the baryon asymmetry
depend exclusively on  the low energy phases in the PMNS matrix. 
Consequently, 
leptogenesis  is maximally connected
to the low energy leptonic CP-violation.  
This conclusion  is clear 
from the expression of the flavour CP asymmetries in terms of a  real $R$ 
matrix, 
$\epsilon_{\alpha}\propto \sum_{\beta, \rho>\beta}\sqrt{m_\beta m_\rho}
(m_\rho-m_\beta)R_{1\beta}R_{1\rho}
{\rm Im}\left(U^*_{\alpha\beta}U_{\alpha\rho}\right)$.
Notice that $\epsilon_1=0$ if $R$ is real   and $\epsilon_{\alpha}=0$ 
if $R$ is real and diagonal. 
Once flavour effects are taken into account,
a   baryon asymmetry is generically generated  from nonzero phases in the
PMNS matrix. 

To illustrate better this point, we provide two examples where the baryon
asymmetry is generated uniquely by the CP phases in the
PMNS matrix.
We will consider the range of values 
$\left(10^9 \lsim  M_1\lsim 10^{12}\right)$ GeV,
for which it is sufficient to consider 
$\epsilon_\tau$, being $\epsilon_2 = - \epsilon_\tau$.
In the first example, we consider the NH spectrum. In the limit 
$M_1\ll M_2\ll M_3$, we obtain  
\begin{eqnarray}
\epsilon_{\tau}&\simeq& \frac{3 M_1}{16\pi v^2}
\frac{(\deltasol \deltaatm)^{1/4} R_{12} R_{13}}{\sqrt{\deltasol/\deltaatm} 
R_{12}^2 + R_{13}^2}c_{13} \nonumber \\
&& \hspace{-1truecm}
\times \left( \frac{1}{2} c_{12} \sin {2 \theta_{23}} 
\sin \frac{\alpha_{32}}{2} - s_{12} c_{23}^2 s_{13} 
\sin\left(\delta - \frac{\alpha_{32}}{2} \right) \right)\, ,
\label{CP-flavour-NH}
\end{eqnarray}
where $c_{ij}=\cos\theta_{ij}$ and $s_{ij}=\sin\theta_{ij}$.
Only the Majorana phase $\alpha_{32}=\alpha_{31} - \alpha_{21}$ plays a role
being the contribution of $m_1$ negligible.
With these expressions, it is straighforward to compute the 
final baryon asymmetry solving the flavoured Boltzmann equations
of Ref. \cite{davidsonetal2}. 
In the IH case, a similar expression holds
for $\epsilon_{\tau}$, but is suppressed for real $R$ with respect
to the one in the NH case by a factor $\sim (\deltasol/\deltaatm)^{3/4}$, 
leading generically 
to an asymmetry which is small. A  sufficiently large  asymmetry can be
recovered in the case of purely imaginary product $R_{11}R_{12}$
or in the supersymmetric version of leptogenesis \cite{appear}.
In the expression (\ref{CP-flavour-NH}) 
the dominant contribution comes from the Majorana CP-violating phase,
while the effects due to $\delta$ are suppressed by $\sin \theta_{13}$.
The Majorana phase $\alpha _{32}$ appears in the expression
   for the effective Majorana mass $\langle m_\nu\rangle$. 
The baryon asymmetry 
depends also  on the combination $\sin \theta_{13} \sin \delta$, 
which enters in the CP-asymmetry
measurable
in future long baseline oscillation experiments.


\begin{figure}[h]
  \centerline{
{\includegraphics[width=6truecm,height=3.5cm]{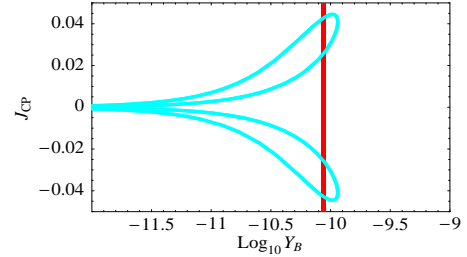}}}
  \caption{The invariant $J_{\rm CP}$ versus the 
baryon asymmetry varying (in blue) $\delta = [0,2\pi]$ in the case of 
hierarchical RH neutrinos and NH  light neutrino mass spectrum   for 
$s_{13}=0.2$, 
$\alpha_{32}=0$, $R_{12} =0.86$,
$R_{13} =0.5$ and  
$M_1 = 5\times 10^{11}$ GeV . The red region denotes the 
$2\sigma$ range for the baryon asymmetry.
}
\label{normal2YBJalpha0s130.2.eps}
\end{figure}
\begin{figure}
  \centerline{
{\includegraphics[width=6truecm,height=3.5cm]{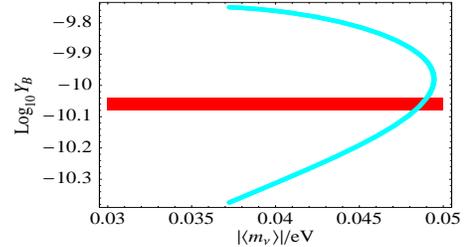}}}
  \caption{The baryon asymmetry $|Y_{\rm B}|$ versus the effective
Majorana mass in neutrinoless double beta decay,
$\meff$, in the case of Majorana CP-violation, 
hierarchical RH neutrinos and IH light neutrino mass spectrum,  
for $\delta = 0$, $s_{13} = 0$, purely imaginary
$R_{11}R_{12}$, 
$|R_{11}| = 1.05$ and $M_1 = 2\times 10^{11}$ GeV.
The Majorana phase $\alpha_{21}$ is varied in the interval
$[-\pi/2, \pi/2]$. 
}
\label{IHneutrinoless.eps}
\end{figure}
We consider the tri-bimaximal mixing case and take
$c_{23}=s_{23}
=1/\sqrt{2}$, 
$s_{12}=1/\sqrt{3}$.  
In Fig. \ref{normal2YBJalpha0s130.2.eps} we show the 
correlation between the baryon asymmetry and the CP invariant 
$J_{\rm CP}$
for a given choice of the parameters and varying the Dirac phase $\delta$. 
Most   values of $J_{\rm CP}$ consistent with the observed baryon 
asymmetry lie well within the sensitivity reachable by superbeam and betabeam 
experiments and  future neutrino
factory. 
In Fig. \ref{IHneutrinoless.eps} we show the correlation
between $Y_{\cal B}$ and $\meff$ in the case of IH light neutrino
mass spectrum and purely imaginary product $R_{11}R_{12}$ (see ref. 
\cite{appear} for details). 

\begin{figure}
  \centerline{
{\includegraphics[width=6truecm,height=3.5cm]{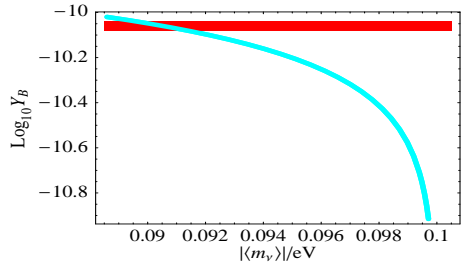}}}
  \caption{ The quantity $\left|\langle m_\nu\rangle \right|$ versus
the baryon asymmetry varying $\alpha_{32}$ between 0 and $\pi/3$  for
the case of degenerate RH neutrinos and QD for light neutrinos for
 $\delta=\pi/3$,
$s_{13}=0.01$,
$M_1 =   10^{10}$ GeV and $m=0.1$ eV.}
\label{RHdegeneratelightQDBYmnu.eps}
\end{figure}

The second example we discuss  is for QD neutrinos.  
To avoid excess of fine-tuning, we choose
quasi-degenerate RH neutrino masses as well, $M_1\sim M_2\sim M_3$; 
 all RH neutrinos  contribute to the baryon asymmetry. 
The  washing out of a given flavour is parametrized 
by $\widetilde{m_{\alpha}}=
\sum_{j}\left|\lambda_{j\alpha}\right|^2 v/M_1$. For
$R$ real, it  
is approximately the same
for all flavours, $\widetilde{m}_{\alpha}\sim m$.  Again,  
for $(10^9\lsim M_1\lsim 10^{12})$ GeV and $R$ real, 
$\epsilon_2=-\epsilon_\tau$.
If
we consider the case in which $M_1\simeq M_2\lsim M_3$, the total CP asymmetry
in the third flavour $\epsilon_\tau$ 
is resonantly enhanced when the decay rate $\Gamma_{N_2}\sim
(M_2-M_1)$ and \cite{appear}
\begin{eqnarray}
\epsilon_\tau &\simeq &  \frac{1}{2 m^2}
\left(\Delta m^2_\odot R_{11}R_{21}-
\Delta m^2_\oplus R_{13}R_{23}\right)\nonumber\\
&\times &\sum_{\rho>\beta}\left(R_{1\rho}R_{2\beta}-R_{1\beta}R_{2\rho}\right)
{\rm Im}\left(U_{3\beta}U_{3\rho}^*\right)\, .
\end{eqnarray}
We may write the matrix $R$ under the form $R=e^{A}$, where $A$ is a real  
matrix satisfying $A^T=-A$. 
In Fig. \ref{RHdegeneratelightQDBYmnu.eps}, 
we show the correlation of the baryon 
asymmetry with the effective Majorana mass 
in neutrinoless double beta  decay. 
A number of projects aim to reach a sentivity to $\left|\langle
m_\nu\rangle\right|\sim\left(0.01-0.05\right)$ eV \cite{aa} and can certainly
probe the region of values of $\left|\langle
m_\nu\rangle\right|$ for successfull  baryon asymmetry 
from the PMNS phases only. 
\noindent
In particular, a direct information
on the Majorana phase $\alpha_{21}$ may come from the  measurement
of $\langle
m_\nu\rangle$, $m$, and $
\sin^2(\alpha_{21}/2)\simeq \left(1-(\left|\langle
m_\nu\rangle\right|^2/m^2)\right)(1/\sin^2 2\theta_{12})$
and might tell us if enough baryon asymmetry may be generated uniquely from
the PMNS phases.
 
Our examples show that the  observation of effects of the CP-violating phases
of $U$
in  neutrino oscillations and/or in the neutrinoless double beta decay
would generically ensure a nonvanishing baryon asymmetry through leptogenesis. 
We will present a more detailed analysis, including the
supersymmetric generalization,   in a forthcoming publication
\cite{appear}.

\noindent
We thank S. Davidson for useful comments.

\end{document}